Article type: Full Paper

# Spontaneous Repairing Liquid Metal/Si Nanocomposite as a Smart Conductive-Additive-Free Anode for Lithium-ion Battery

*By Bing Han[1], Yu Yang[1], Xiaobo Shi[1], Guangzhao Zhang[1, 2], Lu Gong[3], Dongwei Xu[1], Hongbo Zeng[3], Chaoyang Wang[2], Meng Gu[1], Yonghong Deng[1],\**

Dr. Bing Han, Dr. Yu Yang, Dr. Xiaobo Shi, Dr. Guangzhao Zhang, Dr. Dongwei Xu, Prof. Meng Gu, Prof. Yonghong Deng
Department of Materials Science & Engineering, Southern University of Science and Technology of China, Shenzhen 518055, China
E-mail: yhdeng08@163.com (Y. H. Deng)

Dr. Guangzhao Zhang, Prof. Chaoyang Wang
Research Institute of Materials Science, South China University of Technology, Guangzhou 510640, China

Mr. Lu Gong, Prof. Hongbo Zeng,
Department of Chemical and Materials Engineering, University of Alberta, Edmonton, AB T6G 1H9, Canada

**Abstract**

Silicon is a promising candidate for negative electrodes due to its high theoretical specific capacity (~3579 mAh g$^{-1}$) and low lithiation potential (~0.40 V vs Li). However, its practical applications in battery have been inhibited by the large volume change (~400%) induced by Li$^+$-insertion into Si lattices. Here, we attempt to resolve this issue at a fundamental level, and report for the first time a novel liquid metal (LM)-mediated spontaneous repairing conductive-additive-free Si anode for Li-ion battery. The fluidity of LM ensures the eternal contact between Si and the conducting-network during its repeated electrochemical reactions. The as-prepared nano-composite of LM/Si leads to superior performances as characterized by high capacity utilization (2300 mAh g$^{-1}$ at 500 mA g$^{-1}$), long-term stability (968 mAh g$^{-1}$ after 1500 charge-discharge cycles at 8 A g$^{-1}$ with 81.3% retention), high rate capability (360 mAh g$^{-1}$ at 20 A g$^{-1}$, equivalence of 55 C, or full charge/discharge in 65 seconds), and, in particular, an extra-ordinarily high initial coulombic efficiency (95.92%), which is not only the highest reported for Si to the best of our knowledge, but also higher than the mature graphitic carbon anodes. The unique approach described in this work not only resolves the basic stress challenges faced by the promising but often problematic alloy-type materials; in broader context it also provides a universal inspiration to all electrode materials whose electric properties suffer from extreme mechanic upheavals induced by the electrochemical strains during the cell reactions.

**1. Introduction**

The ever increasing demand for new battery chemistries of higher energy and power densities and longer cycle-life presents unprecedented challenges for the intercalation-type materials used in the state-of-the-art lithium-ion batteries (LIBs).[1] Graphitic carbon has been the main anode host since the birth of LIB, and numerous efforts have been made to replace it with an alloy-type host that can provide much higher specific capacities, among which Si is apparently the most promising due to its capability of accommodating ~4 Li$^+$ per Si (3579 mAh g$^{-1}$ for Li$_{15}$Si$_4$ vs. 372 mAh g$^{-1}$ of LiC$_6$).[2,3] However, the extreme volume expansion/contraction (420%) of Si during repeated lithiation/de-lithiation leads to its extensive fracturing and fragmentation, followed by the sustained loss of electrical contact with the bulk electrode and incessant consumption of electrolytes.[4] In the past decade, enthusiastic efforts have been made in attempt to relieve Si from the failure caused by its volume change, which includes the design of various nano-structures so that the extra volume of Si could be accommodated at a nano-scale,[5,22] or the application of elastic[6] or conductive[7] polymeric binders which assist to keep the fragmented Si particles within the electric conducting network. Nevertheless, the improvement of cyclic stability and capacity retention is limited because these solutions resorted to an external restraining force without eliminating the fundamental basis for Si irreversibility[8].

To counter the irreversible loss of Si as active material and consumption of electrolyte, a more effective strategy would be to provide a "spontaneous repairing" mechanism for Si,[9,10] thus any destruction of Si local structure would be instantly repaired, and the resultant fragments would be always contained within the conducting network. In fact, several recent reports have demonstrated such possibility,[11,12] as demonstrated by Bao and coworkers,[13] who designed a self-healing polymer (SHP) based on the reversible hydrogen bonding as the binder for silicon anodes. The Si electrodes using SHP have exhibited superior cycling performance and capacity retention due to intact electrical connections among active materials.

To further perfect this "spontaneous repairing" concept, we seek to employ liquid metal (LM) as the conducting media to construct an electric network in Si electrode. LMs are well-established soft conductors for applications in flexible or stretchable electronics,[14-16] whose fluidity ensures its natural and instantaneous tendency to "self-heal" once its integrality is mechanically disrupted. Because of low melting temperature and small Young's modulus,

LMs can also be easily processed into nanoparticles at ambient temperatures,[17,18] which tend to coalesce upon external stimuli, demonstrating excellent self-healing nature at nano-scale.

In this work, for the first time, we report such a novel self-healing nanocomposite as anodes in LIBs using nanoparticles of Si and LM based on gallium-indium-tin (GaInSn) alloy. It was found that the mechanical stress induced by the volume expansion of Si nanoparticles during electrochemical reactions could be well absorbed by LM nanoparticles, whose coalescence will be induced following the breach of native oxide film upon the initial lithiation of Si, thus significantly enhancing both the conductivity and connectivity within the electrode micro-structures. Excellent electrochemical performance was obtained from such nano-composite because of the synergetic collaboration between the spontaneous repairing LM nanoparticles and high capacity Si nanoparticles.

## 2. Results and Discussion

The LM nanoparticles based on GaInSn were prepared as described in a previous report.[19] Because of its high surface tension, ultrasonication with large power (1000W) was applied to divide the bulk GaInSn LM into nanoparticles (Figure S1). It is noted that ethyl 3-mercaptopropionate serves as an indispensable surfactant to prevent the LM nanoparticles from direct coalescence after division. Upon the addition of Si nanoparticles, the composite was heated at about 60-80 $^{\circ}$C under stirring, during which the homogeneous aggregation of Si and LM nanoparticles occurred with the gradual reduction of LM nanoparticle stability due to the slow evaporation of ethyl 3-mercaptopropionate. The STEM images of the resultant LM/Si nanocomposite revealed that Si nanoparticles of ~80 nm[20,21] are bound at the interface of core-shell-like spherical LM nanoparticles (Figure 1a-d). The elemental mapping based on energy dispersive X-ray spectroscopy (EDS) confirmed the above architecture. The element mapping of the sample further verified that the LM nanoparticle present a homogenous distribution of elements (Figure 1f-h) and the element compositions of these LM nanoparticles were the same as for the bulk, with the Ga:In:Sn ratio being 69:21:10 (Figure S2).

Figure 2a shows the cyclic voltammetry (CV) of the LM/Si anode, which was conducted in a coin-cell in the potential range of 3-0.01 V and at a scanning rate of 0.1 mV s$^{-1}$. The high cut-off was used here to ensure complete delithiation during charge, although a lower value of <1.5 V should be used in actual full cells with essentially the same capacity utilization. In the first lithiation process, the region from 1.1 to 0.6 V represents an irreversible reaction corresponding to the formation of a solid-electrolyte-interphase (SEI), which can be easily identified when comparing with the subsequent cycles. An encouragingly high Coulombic efficiency (CE%) of ~95.92% was achieved in the first precycle at 2 A g$^{-1}$ (Figure 2c). To the best of our knowledge, it is the highest value ever reported for Si-based anodes. More interesting, this CE% is even higher than that of the graphitic carbon anode materials used in commercial LIB. In other words, the consumption of electrolyte during the formation of SEI is kept at a minimum, hinting at a very small surface that requires passivation. Electrochemical impedance spectra were performed on three fully de-lithiated electrodes (3.0 V vs. Li) that were cycled to the 3rd, the 200th, and the 1000th cycles (Figure 2b), respectively. The overall cell impedance gradually increases by three-folds during the long-term cycling, indicating that each cycle still creates new surface that induces the formation of new SEI, but this rate is rather slow considering that Si is the anode host here, which breaks upon each lithiation. It can be speculated that, in the charge-discharge process, the expansion of Si nano-particle must exert pressure on the soft LM nanoparticles, breaching the native film of insulating oxide thereon (Ga$_2$O$_3$) (Figure 1a-b) and establishing the intimate electric contacts with the interior bulk of LM nanoparticles. Benefited from the fluidity of LM, the volume expansion of Si will always be well-accommodated, and the electric contacts established will be permanently maintained when the lithiated Si nano-particle shrinks or expands upon repeated de-lithiation/lithiation.

The intimate contact between the two components and the metallic nature of LM ensures high rate charge-transfer, as evidenced by the rate capability of LM/Si anode (Figure 3a) with current densities from 0.5 to 100 A g$^{-1}$ applied, as well as the capacity retention at different current densities from 0.5 to 20 A g$^{-1}$ (Figure 3b). Remarkable capacity utilization of 2300 mA h g$^{-1}$ was obtained at 500 mA g$^{-1}$ (Figure 3b), and at high rate of 2 A g$^{-1}$, a reversible capacity of 1780 mA h g$^{-1}$ could still be maintained. Even when the current density was pushed to extreme values of 4 A g$^{-1}$, 8 A g$^{-1}$, 16 A g$^{-1}$ and 20 A g$^{-1}$, the LM/Si nanocomposite still maintains reversible capacities of 1500 mAh g$^{-1}$, 1100 mAh g$^{-1}$, 620 mAh g$^{-1}$ and 360 mAh g$^{-1}$, respectively, which is unprecedented for alloy type anode materials. Long-term cycling stability was demonstrated in Figure 3c, where for the first 500 galvanostatic cycles different current densities from 2 to 20 A g$^{-1}$ were applied. After 500 cycles, a moderate current density of 2 A g$^{-1}$ was used, which drains a specific capacity of 1440 mAh g$^{-1}$ with 78% capacity retention, with a fading rate of only 0.044% per cycle.

Long-term stability of LM/Si nano-composite anode is also demonstrated in Figure 4a. At 8 A g$^{-1}$, over 81.3% capacity retention can be retained after 1500 charge-discharge cycles, with a fading rate of only 0.013% per cycle (Figure S4). Using SEM, post-mortem analysis on the electrodes recovered from the long-term cycling revealed that the morphologies of the anodes essentially remain intact within 200 cycles (Figure 4b). After 200 cycles, the flexible inner liquid metal wets the expanded Si nanoparticle, and merges with the latter producing a LM/Si electrode (Figure 4c). In this process the contact surface area between the electrode and electrolyte should dramatically decrease, while the LM/Si cluster become visibly bigger reaching a diameter of ~200 nm. This reduction in surface area is especially beneficial because it minimizes the irreversible consumption of electrolytes required for SEI formation on new electrode surfaces, which was reflected by the especially high average CE% during the cycling (>99%, Figure S3). After 1500 cycles, the LM/Si cluster grows even bigger with a diameter of ~500 nm, with further decreased surface area (Figure 4d). No obvious crack and fragmentation can be observed after such long-term cycling.

To further understand the interaction between LM and Si nanoparticles, controlled samples of pure Si nanoparticle anodes and pure LM nanoparticles are fabricated. The cycling retention test and the rate performance of these controlled anodes, as shown in Figure 5a and 5b, respectively, are self-evident that superior performances (capacity utilization, rate capability and cycling stability as shown in Figures 3 and 4) were the results of LM and Si nano-particles interacting with each other. Reference pure Si anodes and pure LM anodes are tested under similar conditions at 4 A g $^{-1}$, 8 A g $^{-1}$ and 16 A g $^{-1}$, respectively.

The synergy between the soft conductor LM and high capacity Si nanoparticles delivers the excellent electrochemical performances in Li-ion cells (Table S1, Figure 5c). The interaction between LM/Si is schematically described in Figure 7 and Figure S9, which mainly consists of two steps: (1) In the initial lithiation process, the Si nanoparticles expands and breaches the LM droplets; and (2) the flexible inner liquid metal yields to and the subsequently wets the expanded Si. The two components subsequently merge, producing a LM/Si nanocomposite. Due to the "liquid" nature of the composite electrode, which intrinsically minimizes its own surfaces, the contact area between the electrode and electrolyte dramatically decreases accordingly (Figure 7, Figure S3), which significantly reduces the amount of electrolyte consumption required for SEI formation and re-formation, and is responsible for the unusually high initial CE% of 95.92% and the >99% CE% in the following cycles. As demonstrated in Figure 1 and Figure 6, the insulating shells of the LM nanoparticles can be easily broken by external force, which can be quantified using atomic force microscopy (AFM) force measurement (Figure 6a). The AFM tip of radius 8 nm was driven to approach the LM surface, which jumps into contact with the LM surface at some critical separation due to their attractive van der Waals interaction. The AFM tip was then further compressed against the LM surface which led to a positive compressive force and deformation of the LM surface. When the applied load was over ~58 nN, the tip-LM surface breakthrough occurs as indicated by an arrow in the red tracing force curve in Figure 6a. In the retraction process, the force curve does not coincide with the approaching curve due to the adhesion between the AFM tip and the LM surface, which caused a so-called adhesion hysteresis behavior. Figure 6b shows the force-distance profiles between a silica sphere (diameter 5 μm) and the LM surface, where adhesion was also measured during the retraction process. The histogram of the adhesion between the silica sphere and LM surface for 100 measurements over 40 different locations is shown in Figure 6c. This strong adhesion significantly improved both electrical contacts between the two and the mechanical integrity of the nanocomposite during the extreme volume changes of Si nano-particles. Together, these effects contribute to the spontaneous repairing and ensure excellent electrochemical performances observed.

To observe the spontaneous repairing effect of LM on cycling, we performed aberration-corrected STEM and EDS analysis on the electrode before and after cycling. The pristine sample before cycling shows a well-dispersed separate LM and Si morphology (Figure 8a). There is no LM layer on the silicon particle. In sharp contrast, most LM particles breaks after cycling and the LM covers the surface of the cycled silicon aggregation as shown by the overlaid Ga/Si EDS map in Figure 8b. As also found by STEM and EDS analysis in a much larger viewing area, the neat Si electrode has cracks as shown by Figure 8b. However, overlaid Ga/Si maps in Figure 8b show that the LM heals the crack/gap of the silicon electrode as pointed out by the red rectangle. The aberration-corrected STEM results clearly demonstrate that the unique malleability of LM allows the LM to flow into the gap or crack in the silicon electrode. Therefore, the spontaneous repairing is achieved through the ultra-high malleability of LM, which help the electrode to maintain well interconnected morphology and thus a very good capacity retention after long cycles.

## 3. Conclusions

In conclusion, for the first time, we have designed a conductive-additive-free nano-composite anode consisting of LM and Si nano-particles, in which the two components collaboratively generate the excellent electrochemical and mechanical performances, including the long-term stability that rivals the best of Si materials and an initial CE% that even outperforms state-of-the-art graphitic carbon. The unique combination of LM and Si appears to provide an ideal solution to the volume variation issue of alloy-type materials, and possibly provide a universal inspiration to resolve the extreme mechanic upheavals experienced by most alloy-type materials during electrochemical reactions.

## 4. Experimental Section

*Fabrication of GaInSn nanoparticles and composite anode material.* GaInSn bulk samples were prepared with a 7:2:1 weight ratio of Ga, In, and Sn, and GaInSn nanoparticles were fabricated using a sonicator. General procedure was described as follows: 0.5 g GaInSn alloy and 2 mg ethyl 3-mercaptopropionate were added to the solution of 100 mL isopropanol (99%, Kermel, China) under continuous ultrasonication, whose power was controlled at 90% of the maximum (1200 W) for 60 min. The temperature of the sample during ultrasonication was controlled by a 25 ℃ water bath. After sonication, 0.5 g nano-Si particles (average size: 80 nm) were added to the GaInSn nanoparticle isopropanol solution and then sonicated for 20 min. Finally, the composite anode material solution was heated in a temperature range of 60~80 ℃ under continuous stirring (1000 r min$^{-1}$). The residue powder was scraped down for later use after the complete evaporation of isopropanol. The content of Si nanoparticle of the final composite anode material was 50 wt% while the residue part was GaInSn nanoparticles.

*Preparation of LM-Si electrodes.* A typical method was used to fabricate the electrode for electrochemical measurements. The general procedure is presented as following: The powder of composite anode material (180 mg) and Poly (acrylic acid) (PAA, 20 mg) was added to 1 mL isopropanol under continuous stirring (400 r min$^{-1}$) for 48 h. The slurry was then homogeneously coated on Cu foil with blade coater (Laurell, USA). The final mass loading of the electrode after drying completely was 0.28-0.45 mg cm$^{-2}$. Electrodes for the control experiments were fabricated with the pure LM nanoparticles and pure Si nanoparticles using similar procedure.

*Performance measurements.* The morphologies of samples were characterized by STEM (FEI Themis G2 microscope at 300kV), a commercial SEM system (FE-SEM, Tescan mira3, Oxford, Czech Republic) and EDS (FE-SEM, Tescan mira3, Oxford, Czech Republic). EIS and CV of samples were measured by Solarton analytical electrochemical workstation (mode 1470E, England). Charge/Discharge performances of samples were measured by Land charge/discharge instrument (Wuhan Rambo Testing Equipment Co., Ltd.). Then, 1 M LiPF6 dissolved in Ethylene carbonate, Ethyl methyl carbonate and dimethyl carbonate (EC: EMC: DMC = 1:1:1 vol%) were obtained from Shenzhen Capchem Technology Co. Ltd. as electrolyte.

*Force measurements using an atomic force microscope (AFM).* The interaction forces between an AFM tip or a silica colloidal probe and the liquid metal surface in air were measured using a MPF-3D AFM (Asylum Research, Santa Barbara, CA, USA). The silicon nitride AFM tip (NCHV-A AFM probe) was purchased from Bruker (Santa Barbara, CA,

USA) with a spring constant of 42 N m$^{-1}$ and tip radius of 8 nm. The silica colloidal probe was prepared using two-component epoxy glue (EP2LV, MasterBond) to attach a silica microsphere (diameter 5 μm, Sigma-Aldrich) onto a tipless NP-O10 AFM cantilever (Bruker, Santa Barbara, CA, USA). The spring constants of the two AFM probes (tip and the silica sphere) were further calibrated using the Hutter-Bechhoefer thermal fluctuations method.

Prior to force experiments, a relatively large drop of liquid metal was spread on a cleaned glass surface to make a LM surface with a dimension of several millimeters. During a typical force measurement, the AFM tip or silica colloidal probe was driven to approach and then retract from the LM surface at a constant velocity 1 μm s$^{-1}$. The deflection of the cantilever was detected through a laser beam that was reflected from the cantilever into a split photodiode detector, which was further converted to force using the spring constant and Hooke's law. Force measurements were conducted at different locations of the liquid metal, and the adhesion distribution was fitted using Gaussian Model..

**Supporting Information**

Supporting Information is available from the Wiley Online Library or from the author.

**Acknowledgements**

The authors are very much grateful to the funds from the Natural Science Foundation of Guangdong Province, China (grant no. x), the Fundamental Research Foundation of Shenzhen (grant no. x).

Received: ((will be filled in by the editorial staff))
Revised: ((will be filled in by the editorial staff))
Published online: ((will be filled in by the editorial staff))

[1] Armand, M., Tarascon, J. M. (2008). Building better batteries. Nature *451*, 652-657.

[2] Yao, D. H., Yang, Y., Deng, Y. H., Wang, C. Y. (2018). Flexible polyimides through one-pot synthesis as water-soluble binders for silicon anodes in lithium ion batteries. Journal of Power Sources *379*, 26-32.

[3] Gao, Y., Yi, R., Li, Y. C., Song, J. X., Chen, S. R., Huang, Q. Q., Mallouk, T. E., Wang, D. H. (2017). General Method of Manipulating Formation, Composition, and Morphology of Solid-Electrolyte Interphases for Stable Li-Alloy Anodes. J. Am. Chem. Soc. *139*, 17359-17367.

[4] Park, S. J., Zhao, H., Ai, G., Wang, C., Song, X. Y., Yuca, N., Battaglia, V. S., Yang, W. L., Liu, G. (2015). Side-Chain Conducting and Phase-Separated Polymeric Binders for High-Performance Silicon Anodes in Lithium-Ion Batteries. J. Am. Chem. Soc. *137*, 2565-2571.

[5] Chan C. K., Peng H., Liu G., McIlwrath K., Zhang X. F., Huggins R. A., Cui Y. (2008). High Performance Lithium Battery Anodes Using Silicon Nanowires. Nat. Nanotech. *3*, 31-35.

[6] Chevrier, V. L., Liu, L., Le, D. B., Lund, J., Molla, B., Reimer, K., Krause, L. J., Jensen, L. D., Figgemeier, E., Eberman, K. W. (2014) Evaluating Si-Based Materials for Li-Ion Batteries in Commercially Relevant Negative Electrodes. J. Electrochem. Soc. *161*, A783-A791.

[7] Liu, G., Xun, S. D., Vukmirovic, N., Song, X. Y., Olalde-Velasco, P., Zheng, H. H., Battaglia, V. S., Wang, L. W., Yang, W. L. (2011) Polymers with Tailored Electronic Structure for High Capacity Lithium Battery Electrodes. Adv. Mater. *23*, 4679-4683.

[8] Chen, S. Q., Bao, P. T., Huang, X. D., Sun, B., Wang, G. X. (2014). Hierarchical 3D mesoporous silicon@graphene nanoarchitectures for lithium ion batteries with superior performance. Nano Res. *7*, 85-94.

[9] Zhang, G. Z., Chen, Y. H., Deng, Y. H., Ngai, T., Wang, C. Y. (2017). Dynamic Supramolecular Hydrogels: Regulating Hydrogel Properties through Self-Complementary Quadruple Hydrogen Bonds and Thermo-Switch. ACS Macro Lett. *6*, 641-646.

[10] Zhang, G. Z., Ngai, T., Deng, Y. H., Wang, C. Y. (2016). An Injectable Hydrogel with Excellent Self-Healing Property Based on Quadruple Hydrogen Bonding. Macromol. Chem. Phys. *217*, 2172-2181.

[11] Sun, Y., Lopez, J., Lee, H., Liu, N., Zheng, G., Wu, C., Sun, J., Liu, W., Chung, J. W., Bao, Z., Cui, Y. (2016). A Stretchable Graphitic Carbon/Si Anode Enabled by Conformal Coating of a Self-Healing Elastic Polymer. Adv. Mater. *28*, 2455-2461.

[12] Deshpande, R.D., Li, J., Cheng, Y., Verbrugge, M. W. (2011). Liquid Metal Alloys as Self-Healing Negative Electrodes for Lithium Ion Batteries. J. Electrochem. Soc. *158*, A845-A849.

[13] Wang, C., Wu, H., Chen, Z., McDowell, M. T., Cui, Y., Bao, Z. N. (2013). Self-healing chemistry enables the stable operation of silicon microparticle anodes for high-energy lithium-ion batteries. Nat. Chem. *5*, 1042-1048.

[14] Liu, Z. Y., Wang, X. T., Qi, D. P., Xu, C., Yu, J. C., Liu, Y. Q., Jiang, Y., Liedberg, B., Chen, X. D. (2017). High-Adhesion Stretchable Electrodes Based on Nanopile Interlocking. Adv. Mater. *29*, 1603382.

[15] Kim, M-G., Alrowais, H., Pavlidis, S., Brand, Q. (2017). Size-Scalable and High-Density Liquid-Metal-Based Soft Electronic Passive Components and Circuits Using Soft Lithography. Adv. Funct. Mater. *27*, 1604466.

[16] Hirsch, A., Michaud, H. O., Gerratt, A. P., Mulatier, S. D., Lacour, S. P. (2017). Intrinsically Stretchable Biphasic (Solid–Liquid) Thin Metal Films. Adv. Mater. *28,* 4507-4512.

[17] Wu, Y. P., Huang, L., Huang, X. K., Guo, X. R., Liu, D., Zheng, D., Zhang, X. L., Ren, R., Qu, D. Y., Chen, J. H. (2017). A room-temperature liquid metal-based self-healing

anode for lithium-ion batteries with an ultra-long cycle life. Energy Environ. Sci., *10*, 1854-1861.

[18] Ren, L., Zhuang, J. C., Casillas, G., Feng, H. F., Liu, Y. Q., Xu, X., Liu, Y. D., Chen, J., Du, Y., Jiang, L., Dou, S. X. (2016). Nanodroplets for Stretchable Superconducting Circuits. Adv. Funct. Mater. *26*, 8111-8118.

[19] Hu, L., Wang, L., Ding, Y. J., Zhan, S. H., Liu, J. (2016). Manipulation of Liquid Metals on a Graphite Surface. Adv. Mater. *28*, 9210-9217.

[20] Liu, X. H., Zhong, L., Huang, S., Mao, S. X., Zhu, T., & Huang, J. Y. (2012). Size-dependent fracture of silicon nanoparticles during lithiation. Acs Nano, *6*, 1522-1531.

[21] Gu, M., Li, Y., Li, X., Hu, S., Zhang, X., Xu, W., ... & Wang, C. (2012). In situ TEM study of lithiation behavior of silicon nanoparticles attached to and embedded in a carbon matrix. Acs Nano, *6,* 8439-8447.

[22] Ngo, D. T., Le, H. T., Kim, C., Lee, J. Y., Fisher, J. G., Kim, I. D., & Park, C. J. (2015). Mass-scalable synthesis of 3D porous germanium–carbon composite particles as an ultra-high rate anode for lithium ion batteries. Energy Environ. Sci., *8*, 3577-3588.

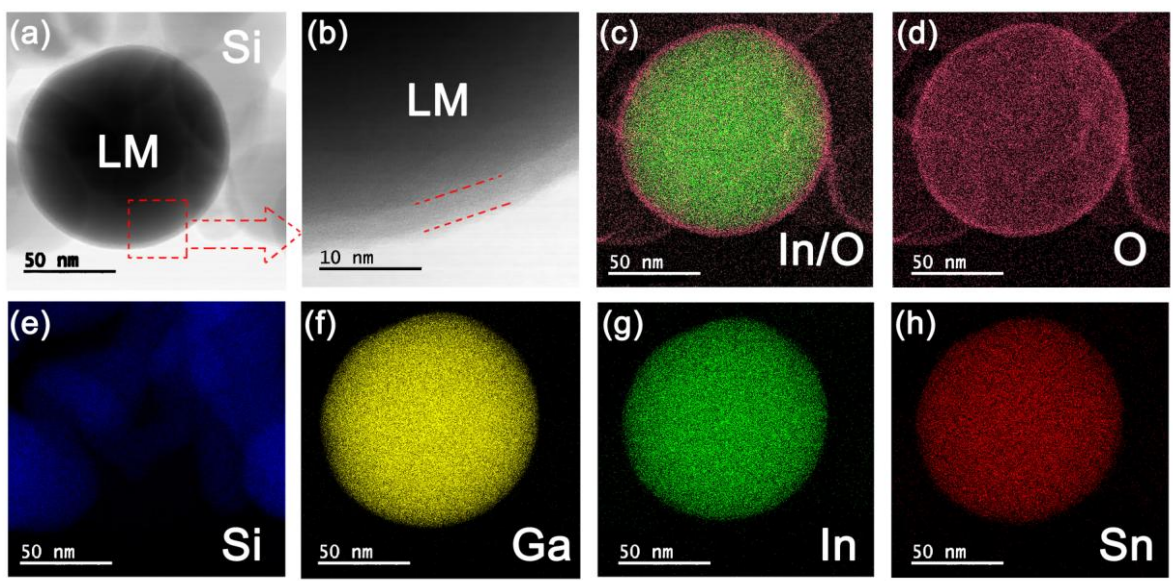

**Figure 1.** STEM characterization of the LM/Si composite at room temperature, as along with element mapping of the same LM/Si composite: (a) Representative bright field STEM image of the LM/Si composites. (b) An enlarging image of the LM nanoparticle in the same LM/Si composite, high resolution bright field STEM image demonstrates the core-shell structure of LM nanoparticle. the black core is the liquid metal (GaInSn alloy), and the lighter part is the coating shell. (c) An element mapping of the LM nanoparticle in the same LM/Si composite. It demonstrates the core-shell structure of LM nanoparticle; the green core is the In metal (GaInSn alloy), and the shell part is the oxygen element. (d) only oxygen element mapping of the same LM nanoparticle. (e)-(h) The element mapping of the same LM/Si nanoparticle. From left to right, the images show LM/Si nanoparticle mapped for Si (blue), Ga (yellow), In (green), and Sn (red).

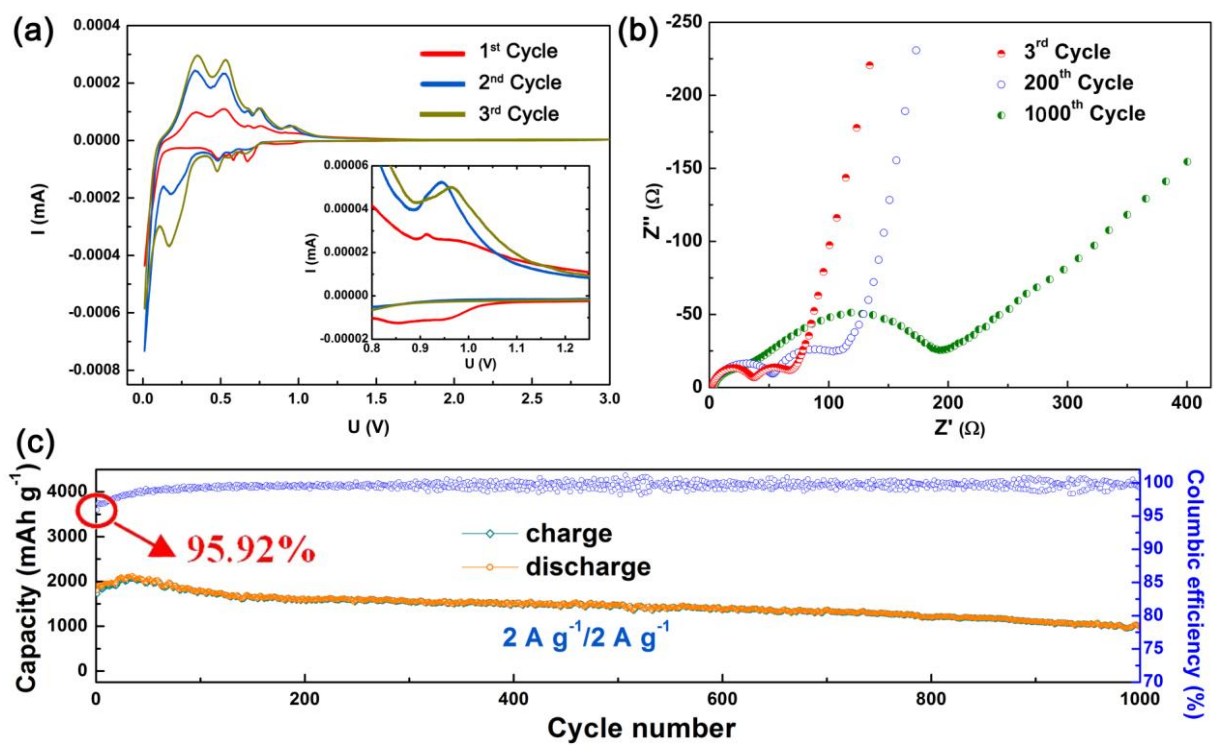

**Figure 2.** Electrochemical performance of the LM/Si nanocomposite. (a) Cyclic voltammetry (CV) plots of the as-made LM/Si anodes scanned at 0.1 mV s$^{-1}$ (b) Nyquist plot of the electrode at different cycles. (c) Long-term cycling behavior of a LM/Si anode cycled at 2 A g$^{-1}$ for 1000 cycles.

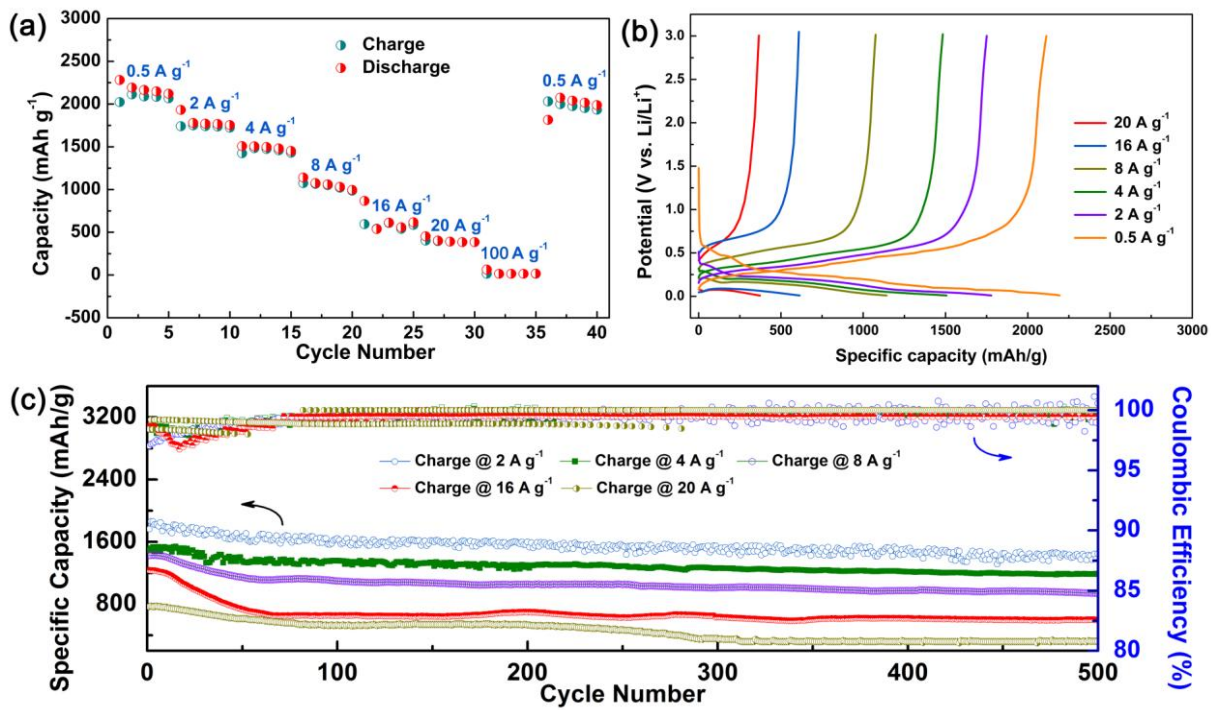

**Figure 3.** Electrochemical performance of the LM/Si nanocomposite. (a) Rate capability and (b) galvanostatic charge-discharge profiles of the as-made LM/Si anode. (c) Long-term cycling behavior of a LM/Si anode cycled using different current density for 500 cycles.

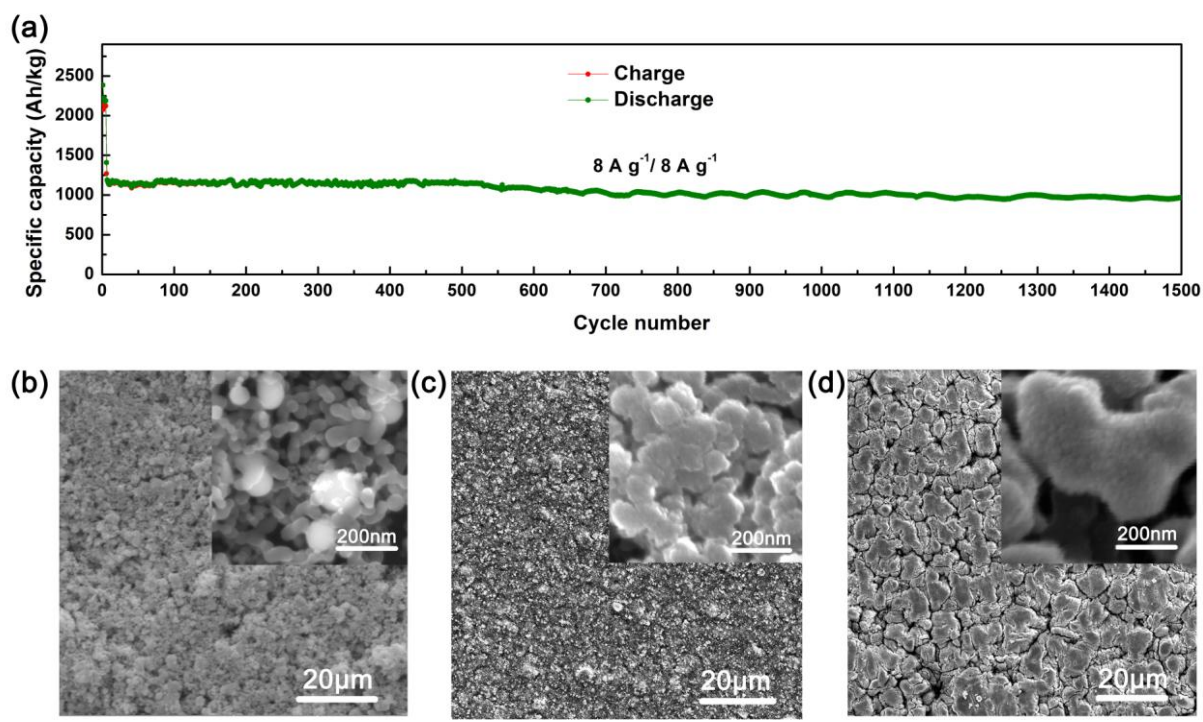

**Figure 4.** Long-term stability of the LM/Si anodes. (a) Cycling of a LM/Si anode under different current densities for 1500 cycles. The morphologies of the recovered LM/Si anodes at (b) initial state, (c) 200th cycle, and (d) 1500th cycle. The Results section should be divided with subheadings.

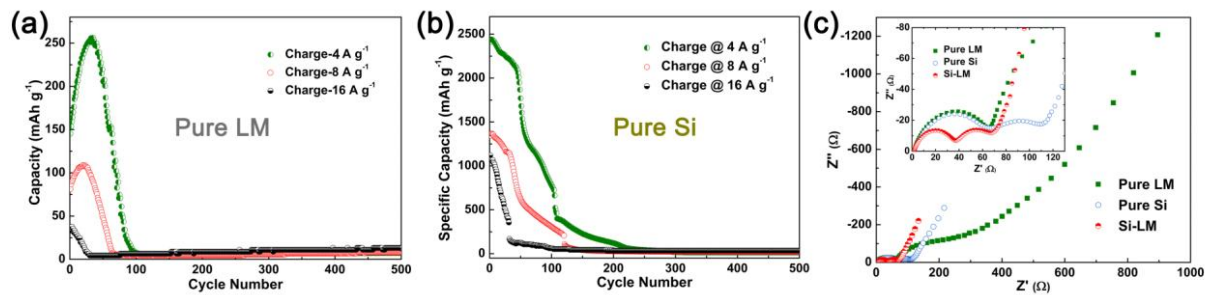

**Figure 5.** Long-term cycling behavior of (a) pure LM anode and (b) pure Si anode cycled using different current density for 500 cycles. (c) Nyquist plot of the different electrode materials (LM/Si, pure LM nanoparticle and pure Si nanoparticle). The inset shows the magnified semicircle at the high-frequency region in corresponding main figures.

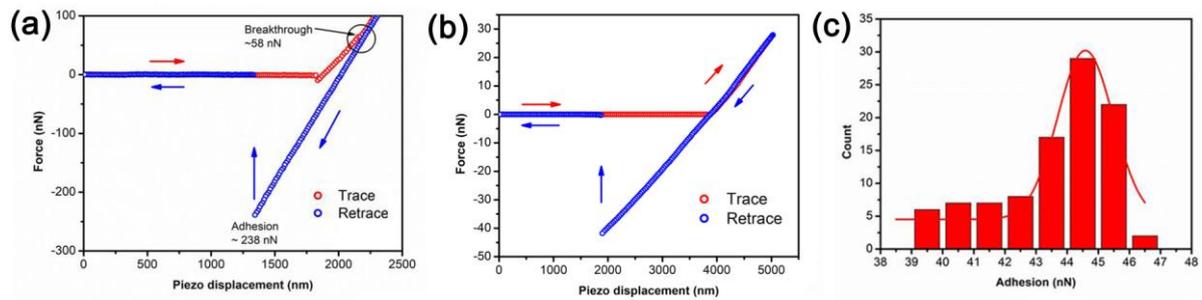

**Figure 6.** (a) Force-distance profiles between an AFM cantilever with tip radius of 8 nm with a LM surface measured using an AFM. (b) Force-distance profiles between a silica sphere (diameter 5 μm) and LM surface. (c) Histogram of the adhesion between the silica sphere and LM surface for 100 measurements over 40 different locations.

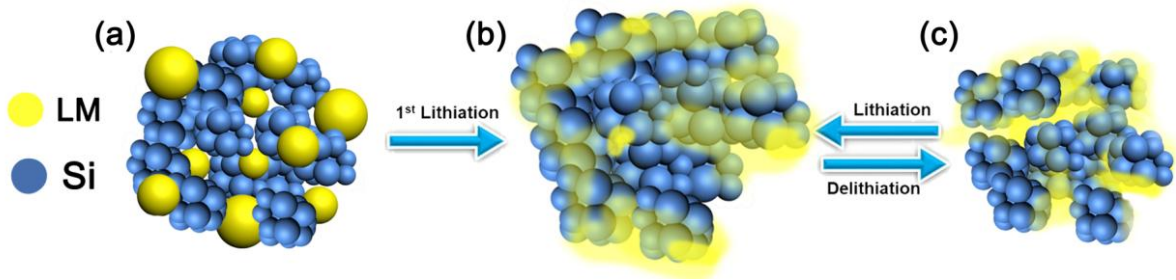

**Figure 7.** Schemes of charge-discharge process of the LM/Si anode.

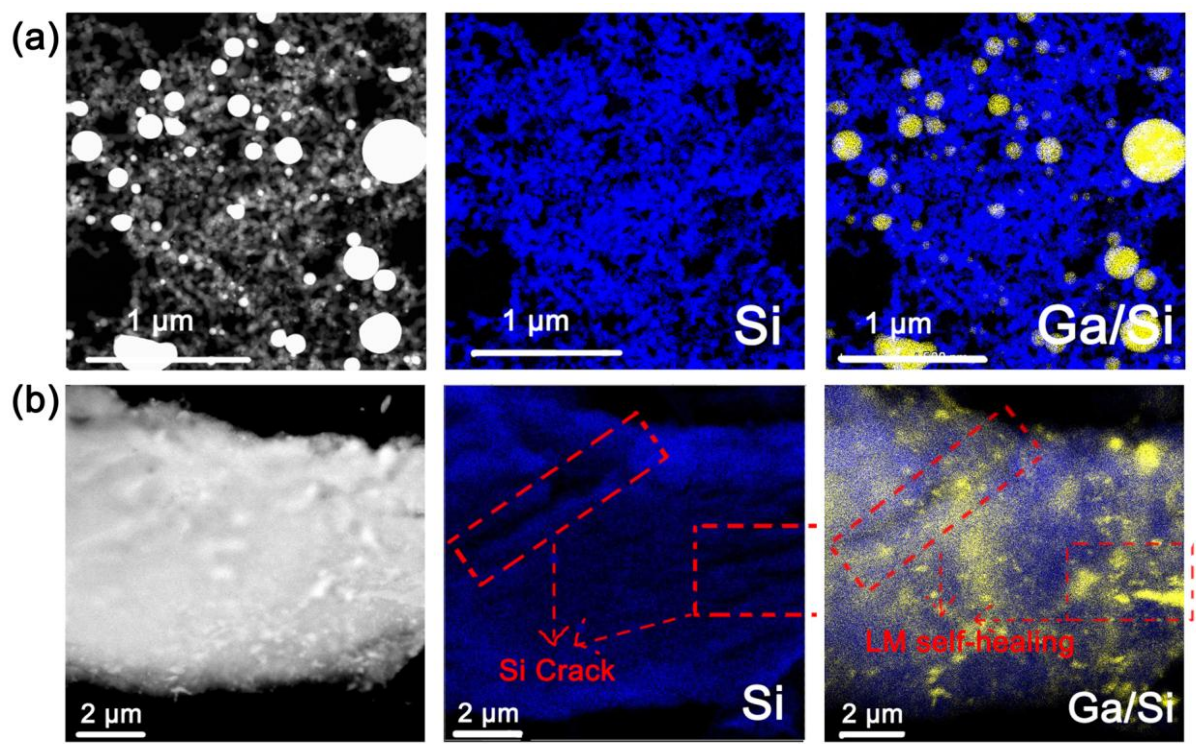

**Figure 8.** A typical STEM image of the LM/Si anode, as along with element mapping of the same LM/Si anode (a) before discharging, and (b) after discharging.

**The table of contents entry**

In this work, we report a smart spontaneous repairing conductive-additive-free anode based on liquid metal (LM)/Si nanocomposite for Li-ion battery, which show excellent electrochemical performances. The unique and facile approach not only enables the promising but problematic alloy-type materials, and also provide a universal inspiration to all materials whose electric properties suffer from extreme mechanic upheavals for practical applications.

**Keywords:** Spontaneous Repairing, Liquid Metal/Si nanocomposite anode, Conductive-Additive-Free , Lithium-ion Battery

**Authors:** *By Bing Han[1], Yu Yang[1], Xiaobo Shi[1], Guangzhao Zhang[1, 2], Lu Gong[3], Dongwei Xu[1], Hongbo Zeng[3], Chaoyang Wang[2], Meng Gu[1], Kang Xu[4], Yonghong Deng[1],\**

**Title:** Spontaneous Repairing Liquid Metal/Si Nanocomposite as a Smart Conductive-Additive-Free Anode for Lithium-ion Battery

**ToC figure**

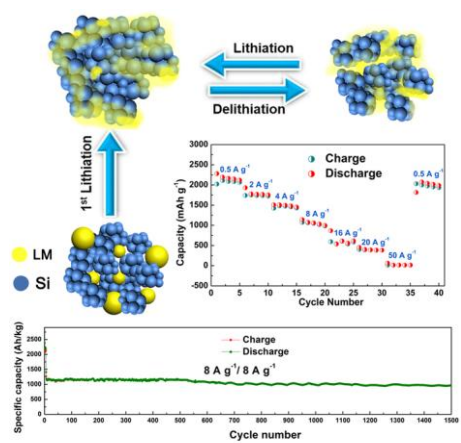